\documentstyle[11pt,newpasp,epsf,twoside]{article}
\def\edcomment#1{\iffalse\marginpar{\raggedright\sl#1\/}\else\relax\fi}
\reversemarginpar

\newcommand{\HI}{H\,{\sc i}}
\newcommand{\HII}{H\,{\sc ii}}
\newcommand{\kms}{km\,s$^{-1}$}
\newcommand{\msun}{M$_{\odot}$}
\newcommand{\mhi}{M$_{\rm HI}$}

\newcommand{\Ha}{H$\alpha$}
\newcommand{\nhi}{\mbox{$N_{\rm HI}$}}
\newcommand{\Lya}{\mbox{Ly$\alpha$}}
\newcommand{\cm}{cm$^{-2}$}

\newcommand{\oiii}{\hbox{[O\,{\sc iii}]}}

\begin{document}
\title{Discovery of Intergalactic \HII\ Regions}
 \author{E. V. Ryan-Weber}
\affil{University of Melbourne}
\author{M. E. Putman}
\affil{CASA, University of Colorado}
\author{K. C. Freeman}
\affil{RSSA, Australian National University}
\author{G. R. Meurer}
\affil{The Johns Hopkins University}
\author{R. L. Webster}
\affil{University of Melbourne}

\begin{abstract}
We have discovered a number of very small isolated \HII\ regions 20-30
kpc from their nearest galaxy. The \HII\ regions appear as tiny
emission line dots (ELdots) in narrow band images obtained by the NOAO
Survey for Ionization in Neutral Gas Galaxies (SINGG). We have
spectroscopic confirmation of 5 isolated \HII\ regions in 3
systems. The \Ha\ luminosities of the \HII\ regions are equivalent to
the ionizing flux of only 1 large or a few small OB stars each. These
stars appear to have formed in situ and represent atypical star
formation in the low density environment of galaxy outskirts. In situ
star formation in the intergalactic medium offers an alternative to
galactic wind models to explain metal enrichment. In interacting
systems (2 out of 3), isolated \HII\ regions could be a starting point
for tidal dwarf galaxies.
\end{abstract}

\section{Introduction}

Isolated \HII\ regions in the extreme outskirts of galaxy halos
(Gerhard et al., 2002) and in gaseous tidal debris (Ryan-Weber et al.,
2003c; Oosterloo et al., 2003) have recently been discovered. Isolated
\HII\ regions indicate the formation of OB stars in atypical
environments. Their existence poses questions about the conditions
required to form stars. Star formation usually occurs in the inner
parts of galaxies and is aided by a high density of gas that is
unstable against gravitational collapse and shielded from the
extragalactic ionizing background.

Here we discuss five small isolated \HII\ regions. The \HII\ regions
were discovered in the NOAO Survey for Ionization in Neutral Gas
Galaxies (SINGG). SINGG is an \Ha\ survey of an \HI-selected sample of
nearby galaxies. The survey is composed of nearly 500 galaxies from
the \HI\ Parkes All-Sky Survey (HIPASS, Meyer et al., 2003), of these
about 300 have been observed in \Ha. The \HII\ regions appear as tiny
Emission Line dots (ELdots) at projected distances up to 30 kpc from
the apparent host galaxy and at least beyond two R$_{25}$ (R-band
isophotal radii with $\mu_R =$ 25 ${\rm mag\, arcsec^{-2}}$). This is
typically much further from the apparent host than outer disk \HII\
regions in spirals (Ferguson, 1998b). Isolated \HII\ regions are
described as `intergalactic' as they lie well beyond the main optical
radius of the nearest galaxy, but may or may not be kinematically
bound to it.

\section{Observations}

Continuum R-band and narrow band \Ha\ images of local gas-rich
galaxies were taken with the CTIO 1.5m telescope as part of SINGG. The
candidate isolated \HII\ regions were identified as unresolved high
equivalent width (EW) sources outside the optical disk of each
galaxy. The isolated \HII\ regions have \Ha\ fluxes in the range
$9.7\times10^{-16}$ to $2.1\times10^{-15}$ erg s$^{-1}$\cm. Assuming
the distance to the isolated \HII\ regions is the same as that of the
host galaxy in each system, the \Ha\ luminosities are in the range
$6.9\times10^{36}$ to $3.5\times10^{38}$ erg s$^{-1}$. In most cases
the isolated \HII\ regions are barely detected in continuum emission
in the SINGG R-band images with a typical 5$\sigma$ detection limit of
$\sim10^{-18}$ erg s$^{-1}$\cm\AA$^{-1}$.

Spectra of isolated \HII\ region candidates were obtained with the
double beam spectrograph (DBS) on the RSAA 2.3m telescope in September
2002. The 5 detected isolated \HII\ regions were confirmed as \Ha\
(6563 \AA) emission line systems, with recessional velocities close to
that of their respective host galaxies. Some isolated \HII\ regions
were also detected in \oiii\ (5007 \AA). The presence of both \Ha\ and
\oiii\ lines places the isolated \HII\ regions at comparable
recessional velocities to the galaxy (or galaxies) in each field, and
rules out the possibility that they are background emission line
systems. In addition to the \Ha\ images and spectra, Australia
Telescope Compact Array (ATCA) \HI\ maps are available for two
systems, NGC 1533 and ESO~149-G003. The ATCA data reduction is
detailed in Ryan-Weber et al. (2003b).

\section{Three Systems with Confirmed Isolated \HII\ Regions}

\begin{figure} 
\plotone{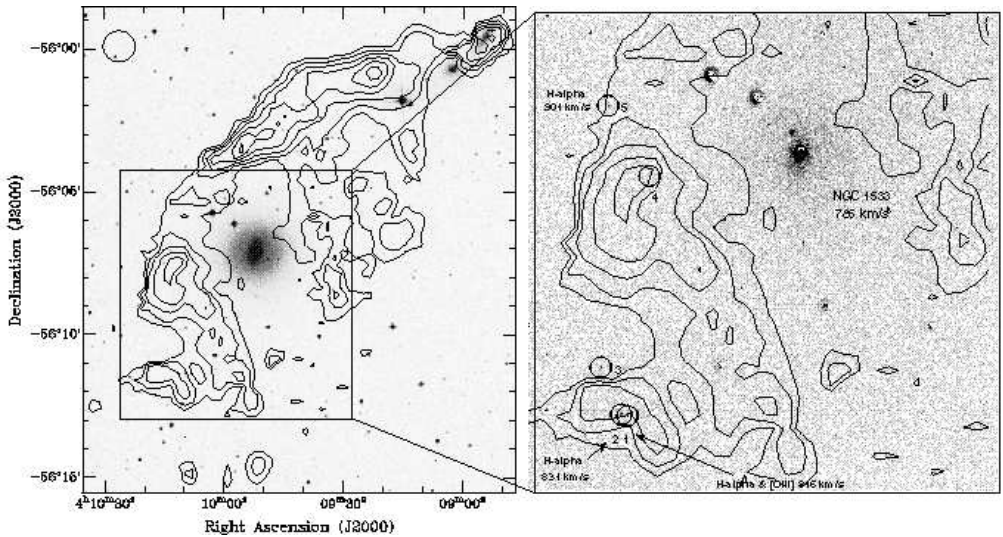}
\caption{NGC 1533: DSS image with ATCA \HI\ contours overlaid at 1.6,
  2.0, 2.4, 2.8, 3.2, 3.6 and 4 $\times10^{20}$ \cm. The beam is given
  in the top left corner. The insert shows the continuum subtracted
  \Ha\ image with the isolated \HII\ regions labelled and \Ha\
  velocities given.}
\end{figure}

\subsection{NGC~1533}

Figure 1 shows a DSS image of NGC~1533 overlaid with
ATCA \HI\ contours. The \HI\ distribution around NGC~1533 consists of
two major arcs, the NW cloud and the SE cloud. No obvious optical
counterpart to this \HI\ feature is seen in the DSS nor R band SINGG
image. The total \HI\ mass of the system is 7$\times10^{9}$~\msun. The
SE cloud contains around one third of this total \HI\ mass. The
projected distance between the \HI\ arcs and the optical centre of
NGC~1533 ranges from 2\arcmin\ to 11.7\arcmin\ (12 and 70 kpc). The
velocities of the three confirmed isolated \HII\ regions compare well
with the velocity of NGC~1533 and lie within the range of \HI\
velocities in the SE cloud (883 \kms\ with a width at 50\% peak,
$w_{50}=71$ \kms). The peculiar distribution of \HI\ is thought to
arise from the destruction of a galaxy to form a tidal remnant around
NGC 1533. N-body/SPH numerical simulations showing the orbital
evolution of a LSB galaxy in NGC~1533's gravitational potential
support this hypothesis Ryan-Weber (2003a).

Are these isolated \HII\ regions progenitors to a tidal dwarf galaxy?
Since the gas and isolated \HII\ regions are bound to the galaxy, it
is likely that the stars formed will also remain bound in the tidal
debris. There is certainly a reservoir of gas from which more stars
could form, so it is possible in this case that a tidal dwarf galaxy
could emerge. Interestingly, the isolated \HII\ regions do not appear
to be correlated with the densest regions of \HI\ and are located in
the SE cloud only. At this resolution ($\sim$6 kpc) the densest region
of \HI\ is the center the NW cloud. The stellar concentration of a
tidal dwarf galaxy is usually located in the densest regions of \HI,
mapped in 21-cm at similar resolutions ($\sim$ 4kpc, e.g. Duc et
al. 2000).

\subsection{HCG~16}
\begin{figure} 
\plotone{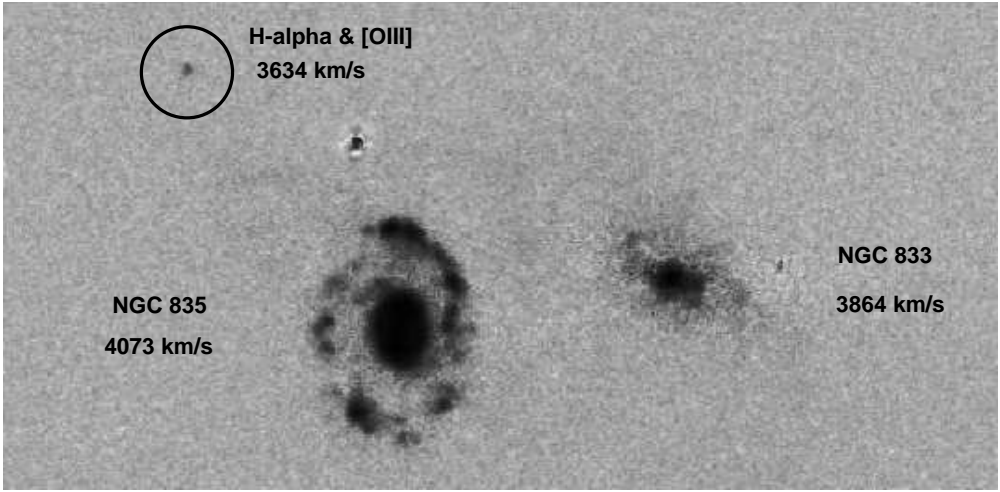}
\caption{HCG 16: Continuum subtracted \Ha\ image with the isolated
  \HII\ region and two members of the galaxy group labelled. The other
  object in the field is residuals of a foreground star.}
\end{figure}

The isolated \HII\ region in the compact group HCG~16, shown in Figure
2, is near the two galaxies NGC~835 and NGC~833. The velocity of the
isolated \HII\ region (3634 \kms) sits on the lower edge of the \HI\
emission measured by HIPASS (velocity = 3917 \kms, $w_{50}=$ 288 \kms)
and below the optical velocities of NGC 835 and NGC 833 (4073 and 3864
\kms\ respectively, from NASA/IPAC Extragalactic Database,
NED). Verdes-Montenegro et al. (2001) have published a VLA map of HCG
16, showing \HI\ in NGC 835 and 833 with a large tidal feature to the
NE (overlapping the isolated \HII\ region position) that joins other
group members several arcminutes away to the east.

\subsection{ESO~149-G003}
\begin{figure} 
\plotone{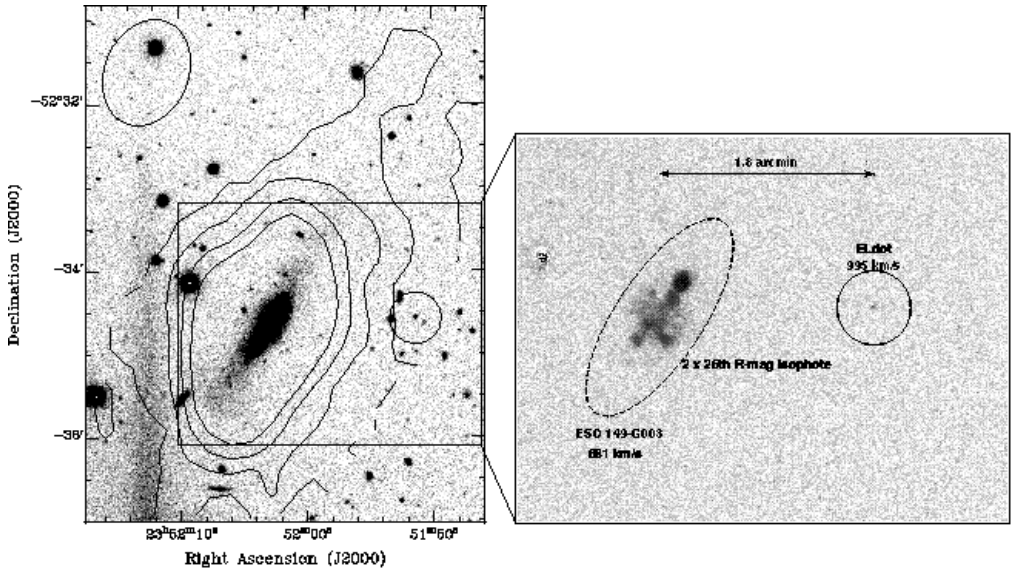}
\caption{ESO~149-G003: R-band image with ATCA \HI\ contours overlaid
  at 0.5, 1.0, 1.5 and 2.0 $\times10^{20}$ \cm. The beam is given in the top
  left corner. The insert shows the continuum subtracted \Ha\ image
  with the isolated \HII\ region and galaxy labelled.}
\end{figure}

The velocity of the isolated \HII\ region near the irregular galaxy
ESO~149-G003 (see Figure 3) is quite offset from its apparent host
galaxy (314 \kms, assuming the single emission line is indeed
\Ha). The narrow ($w_{50}=39$ \kms) \HI\ spectrum of the galaxy shows
no anomalous velocity gas. Follow-up ATCA observations show no \HI\
emission at the velocity and position of the isolated \HII\ region to
a $3\sigma$ limit, assuming a distance of 6.5 (12) Mpc of \mhi\
$=8\times 10^5$ ($3\times 10^6$) \msun. The spatial and kinematic
evidence suggests that the isolated \HII\ region is perhaps a chance
object in the field, rather than an associate of ESO~149-G003. If the
emission line is \Ha, this object holds interesting implications for
the census of intergalactic matter. It could be sub-galactic halo
forming its first generation of stars.  However, since only one
emission line (\Ha) is detected in this case, the possibility that
this candidate isolated \HII\ region is a background emission line
source cannot be ruled out, for example H$\beta$ at z$\sim$0.3 or
\oiii\ at z$\sim$0.2.

\section{Underlying Stellar Population}

The \Ha\ luminosities indicate that each isolated \HII\ region is
illuminated by the equivalent of $0.2-8$ O5V stars each (Vacca et
al. 1996). The least luminous isolated \HII\ region could be ionized
by a single O8 star. The isolated \HII\ regions could be ionized by a
single massive star, or by the massive star or stars in a cluster. The
very low continuum emission does rule out a significant underlying
stellar population and suggests that isolated \HII\ regions are due to
newly formed clusters where no stars existed previously. The low
continuum emission also separates isolated \HII\ regions from \HII\
galaxies and tidal dwarf galaxies. Three of the 5 isolated \HII\
regions have EW(\Ha)$>$1000 \AA. By comparison, \HII\ regions in the
outer arms of spiral galaxies (beyond the B-band 25th-magnitude
isophote) measured by Ferguson (1998a) have an average EW(\Ha)=364
\AA. Furthermore, most of the group of star forming dwarf galaxies in
A1367 have EW(\Ha)$<$100 \AA\ (Sakai et al., 2002).

\section{IGM enrichment}

As they evolve, OB stars increase the metal abundance in their local
environment. Absorption line studies show that the intergalactic
medium (IGM) and galaxy halos, including our own, are enriched (e.g.,
Tripp et al., 2002; Collins et al., 2003). Isolated \HII\ regions
provide a potential source for this enrichment. In situ star formation
in the IGM offers an alternative to galactic wind models to explain
metal enrichment hundreds of kilo-parsecs from the nearest galaxy.

Although the \Ha\ luminosities are small, an estimate of the Star
Formation Rate (SFR) can be obtained by a standard relation
(Kennicutt, 1998). Summing the \Ha\ luminosities from the 5 isolated
\HII\ regions in the NGC~1533 system, the SFR $=3\times10^{-3}$ \msun
yr$^{-1}$. The total yield of metals for a normal stellar cluster is
$\sim$0.025 (Maeder, 1992). Simulations of the dynamical evolution of
\HI\ gas around NGC 1533 show that it could last up to 1$\times10^{9}$
yrs (Ryan-Weber 2003a). This is considered an upper limit since no
consumption of gas due to the formation of stars in taken into
account. If the SFR is maintained for 1$\times10^{9}$ yrs, metals will
pollute the 2.4$\times10^{9}$ \msun\ of \HI\ in the SE cloud,
resulting in a metallicity of $\sim1-2\times10^{-3}$
solar. Alternatively if the SFR was not continuous and corresponded to
a single population of stars only, the resulting metallicity would be
negligible ($\sim4\times10^{-6}$ solar).

How does this compare to the abundances seen in \Lya\ absorption line
systems? \HI\ in the vicinity of the NGC 1533 isolated \HII\ regions
has \nhi\ $=1-4\times10^{20}$ \cm, similar to damped \Lya\ absorption
(DLA, \nhi\ $\geq2\times10^{20}$ \cm). The metallicity of DLA gas at
low redshift varies from 0.01 solar (e.g I Zw 18, Aloisi et al. 2003)
to solar. Depending on the initial metallicity, the isolated \HII\
regions would enrich the NGC 1533 system by 20 percent at the most. At
higher redshifts however, this increase in metallicity could be more
significant. Prochaska et al. (2003) find a DLA `metallicity floor' at
$\sim1.4\times10^{-3}$ solar, over a redshift range from 0.5 to
5. Intergalactic star formation may have contributed to this. In
addition, since collisions and tidal disruptions of galaxies were more
common at higher redshifts, the amount of high \nhi-gas outside
galaxies was greater and therefore the intergalactic star formation
rate could have been higher in the past.

\acknowledgments{Thanks to the SINGG team for the images. This
  research has made use of the NASA/IPAC Extragalactic Database
  (NED). Digitized Sky Survey (DSS) material (UKST/ROE/AAO/STScI) is
  acknowledged.}

\end{document}